\documentclass[runningheads]{llncs}
\usepackage{listings}
\usepackage{amssymb}
\usepackage{amsmath}
\usepackage{pifont}
\usepackage{cite}
\usepackage[T1]{fontenc}
\usepackage{microtype}
\usepackage{ifthen}
\usepackage{graphicx}
\usepackage{comment}
\usepackage[ruled,linesnumbered, noend]{algorithm2e}
\usepackage{siunitx}
\usepackage{placeins} % for \FloatBarrier

\usepackage[dvipsnames]{xcolor}
\usepackage{listings-rust}
\usepackage{lstlinebgrd}
\usepackage{lmodern}
\usepackage{booktabs}
\usepackage{makecell}
\usepackage{tikz}
\usepackage{placeins}
\newcommand{\xmark}{\ding{55}}%
\usepackage[hidelinks]{hyperref}
\hypersetup{
    colorlinks=true,
    linkcolor=RedViolet,
    anchorcolor=black,
    citecolor=blue,
    urlcolor=black,
    breaklinks=true
}
\usepackage{orcidlink}

\usepackage[capitalise]{cleveref}
\Crefname{equation}{}{}
\Crefname{appendix}{Appx.}{Appxs.}
\Crefname{section}{Sect.}{Sects.}
\Crefname{figure}{Fig.}{Figs.}

\usepackage{enumitem}
\usepackage{csquotes}
\usepackage{color}
\newif\ifarxiv \arxivtrue  % set before macro.tex so the default is overridden
\input{macro}
\usetikzlibrary{matrix, shapes.geometric, arrows.meta, positioning}

\newcommand{\toolname}[0]{{\sc VeriStruct}}

\makeatletter
\newlength{\sectionspace}      
\newlength{\subsectionspace}   
\newlength{\subsubsectionspace}
\newlength{\paragraphspace}    
\newcommand{\addperiod}[1]{#1.}
\renewcommand\section{\@startsection{section}{1}{\z@}%
                       {-18\sectionspace \@plus -4\sectionspace \@minus -4\sectionspace}%
                       {12\sectionspace \@plus 4\sectionspace \@minus 4\sectionspace}%
                       {\normalfont\large\bfseries\boldmath
                        \rightskip=\z@ \@plus 8em\pretolerance=10000 }}
\renewcommand\subsection{\@startsection{subsection}{2}{\z@}%
                       {-18\subsectionspace \@plus -4\subsectionspace \@minus -4\subsectionspace}%
                       {8\subsectionspace \@plus 4\subsectionspace \@minus 4\subsectionspace}%
                       {\normalfont\normalsize\bfseries\boldmath
                        \rightskip=\z@ \@plus 8em\pretolerance=10000 }}
\renewcommand\subsubsection{\@startsection{subsubsection}{3}{\z@}%
                       {-18\subsubsectionspace \@plus -4\subsubsectionspace \@minus -4\subsubsectionspace}%
                       {  -0.5em \@plus -0.22em \@minus -0.1em}%
                       {\normalfont\normalsize\bfseries\boldmath\addperiod}}
\renewcommand\paragraph{\@startsection{paragraph}{4}{\z@}%
                       {-12\paragraphspace \@plus -4\paragraphspace \@minus -4\paragraphspace}%
                       {-0.5em \@plus -0.22em \@minus -0.1em}%
                       {\normalfont\normalsize\itshape}}
\makeatother

\newcommand{\genfunc}[0]{\textbf{GenAnnos}}
\newcommand{\fixfunc}[0]{\textbf{RepairAnnos}}

\title{\toolname{}: AI-assisted Automated Verification of Data-Structure Modules in Verus}
\author{Chuyue Sun\inst{1}\orcidlink{0009-0005-9226-3688} \and
Yican Sun\inst{2}\orcidlink{0000-0002-0370-1676}\thanks{Work was accomplished while the author was visiting Stanford University.} \and
Daneshvar Amrollahi\inst{1}\orcidlink{0000-0003-0954-7881} \and
Ethan Zhang\inst{1} \and
Shuvendu Lahiri\inst{3}\orcidlink{0000-0002-4446-4777} \and
Shan Lu\inst{3}\orcidlink{0000-0002-0757-4600} \and
David Dill\inst{1}\orcidlink{0000-0002-6189-0866} \and
Clark Barrett\inst{1}\orcidlink{0000-0002-9522-3084}}
\institute{Stanford University \and
School of Computer Science, Peking University, Beijing, China \and
Microsoft Research}
\newenvironment{talign*}
 {\let\displaystyle\textstyle\csname align*\endcsname}
 {\endalign}

\renewcommand{\paragraph}[1]{\medskip\noindent\textbf{#1.}}

\makeatletter
\@mparswitchfalse%
\makeatother
\normalmarginpar %for right-handed notes and lines, or
\begin{document}

% Mostly about background

\maketitle
\begin{abstract}
%Program verification is essential for ensuring software correctness, a task that has become increasingly critical as AI-generated code is often erroneous or insecure. However, verification is labor-intensive. To alleviate this burden, recent research explores AI-assisted automation but remains limited to verifying an individual function.

We introduce \toolname{}, a novel framework that extends AI-assisted automated verification from single functions to more complex data structure modules in Verus. \toolname{} employs a planner module to orchestrate the systematic generation of abstractions, type invariants, specifications, and proof code. To address the challenge that LLMs often misunderstand Verus' annotation syntax and verification-specific semantics, \toolname{} embeds syntax guidance within prompts and includes a repair stage to automatically correct annotation errors.
%
%We evaluate \toolname{} on eleven benchmark modules from the Verus GitHub repository. \toolname{} successfully verifies ten modules and 106 out of 129 functions (82.2\%) overall—representing a substantial improvement over existing baselines and demonstrating the framework's effectiveness.
%
%the framework incorporates syntax guidance within prompts and adds a repair stage to correct errors in the annotations.
%
In an evaluation on eleven Rust data structure modules,  \toolname{} succeeds on ten of the eleven, successfully verifying 128 out of 129 functions (99.2\%) in total. These results represent an important step toward the goal of automatic AI-assisted formal verification.

    %We present a framework that automatically synthesizes and checks formal specifications for data-structure implementations in Verus, a formal verifier for Rust.
    %Given a Rust module and its unit tests, we defined a novel LLM workflow that verifies entire data-structure modules by synthesizing module-level abstractions--\texttt{View} traits and type invariants--alongside method contracts and proofs. Our work generalizes prior works that target the verification of a single method in isolation.
    %Our work generalizes prior works that target the verification of a single method in isolation.
    %Experimental results show that ..

    %Given a Rust module and its unit tests, a tailored LLM workflow produces the required artifacts. Unlike prior work that targets single methods, our approach verifies entire data-structure modules by synthesizing module-level abstractions—view traits and type invariants—alongside method contracts and proofs. We also add specialized, syntax-aware repair procedures that iteratively fix errors in the generated artifacts.
\end{abstract}

\section{Introduction}
\label{sec:intro}

The power of generative AI introduces new risks to our critical digital infrastructure.  AI systems have a remarkable ability to analyze code, making it much easier for a malicious actor to find and exploit vulnerabilities~\cite{BBB+23}.  At the same time, more and more code is being written by or with the assistance of AI systems.  While it is true that such AI systems can greatly increase the productivity of programmers, they can also introduce additional correctness errors and security vulnerabilities~\cite{insecure1,bug1}.  Thus, productivity gains come at the risk of churning out mountains of buggy, insecure code.

Program verification can mitigate these risks by mathematically proving that code---whether human-written or AI-generated---is free of bugs and vulnerabilities.
However, program verification relies on the addition of sophisticated \emph{logical annotations} to code, including preconditions, postconditions, invariants, and auxiliary proof blocks.  These annotations, which we also refer to as \emph{specifications}, are used to generate verification conditions, which are then proved automatically by automated reasoning tools like satisfiability modulo theories (SMT) solvers~\cite{BSST21}.
The application of program verification in practice has been severely limited because of the enormous effort and expertise required to create such annotations.

Fortunately, the power of generative AI is also shifting the calculus on the practicality of program verification.  Recently, there have been encouraging results on using large language models (LLMs) to \emph{automatically} specify and verify software~\cite{clover,cav24llm,autoverus,classinvgen,dafnysinglemethod}.
Our work aims to build on these efforts with the vision of one day making it possible to have AI-assisted program verification at scale and verified libraries of reusable code~\cite{verilib}.

So far, AI-assisted verification efforts have focused on simple textbook algorithms consisting of a single function.
In this paper, we extend this by developing a novel workflow for the verification of data structure modules. The ability to verify data structures is a crucial advance since: (1) data structures are ubiquitous in everyday programming, forming the foundation of many software systems; and (2) verifying library data structures lowers the burden for verifying more complex code, as client code can then rely on the correctness of these already-proven components.

We implement our workflow in \toolname{}, an AI-assisted automated verification framework for Verus~\cite{verus,verus-practical}.  Verus is a state-of-the-art program verification extension for Rust. We target Verus because it has been successfully applied to the construction of verified systems~\cite{verus-practical}, and its design allows developers to write verification annotations using a syntax closely aligned with Rust. This design not only lowers the barrier for Rust engineers to adopt formal verification but also positions Verus to have an increasing impact as Rust continues to gain traction in mainstream software development. 

The input to \toolname{} consists of 
the Rust source code for the module to be verified and a unit test suite that illustrates the intended usage and expected behavior of the data structure. The test suite serves a dual purpose: it helps ensure that the generated formal specifications align with the developer's intent, and it helps rule out trivial or vacuous specifications~\cite{test-intent}.
%a unit test suite that illustrates the intended usage and expected behavior of the data structure. The test suite serves to ensure that the generated specifications align with developer intent and to rule out trivial or vacuous specifications. %[The unit test can be written by the user, or automatically generated by LLM].
%\end{enumerate}
%
%\paragraph{Output of Our Framework}
\toolname{} automatically augments the provided Rust implementation with annotations needed for program verification. This augmented code, together with the provided unit test suite, are then sent to the Verus verifier, which attempts to formally verify the assertions in the test suite. If the verification check succeeds, the formally verified code is returned to the user.  Otherwise, \toolname{} attempts to extend or repair the annotations based on the feedback from the verifier. %A concrete example of this process is presented below.

%\paragraph{Challenges}
We highlight two challenges for LLM-assisted verification of data-structure modules.
The first challenge is that verifying data structures is inherently more complex than verifying a single function. Specifically, data-structure verification often requires two additional elements:
($i$) a suitable mathematical abstraction that allows the verifier to reason logically about the data structure~\cite{reynolds-seplogic,parkinson-bierman,ohearn-cacm}; and ($ii$)
so-called \emph{type invariants} that must be preserved by operations on the data structure~\cite{classinvgen,pldi-invariant}.
Moreover, data-structure modules usually expose multiple methods that must be verified jointly under a shared type invariant. This stands in contrast to other recent approaches, which focus on generating a single specification artifact (e.g., proof blocks for an individual function or automated synthesis of class invariants)~\cite{autoverus,dafny-annotator,dafnysinglemethod,cav24llm,classinvgen,lemur,loopy}. In short, our verification task extends beyond the scope of these existing techniques.

The second challenge lies in the LLMs' limited understanding of Verus' specialized syntax and verification-specific semantics. For instance, annotations in Verus are only allowed to invoke ``specification functions,'' which cannot modify global state. However, the LLM sometimes suggests invoking a regular Rust function. Such deficiencies primarily stem from the scarcity of Verus code in existing training data.

%the model occasionally invokes executable functions when generating specifications--an operation disallowed in Verus because such functions may modify global states, violating the purity required of logical specifications. This deficiency primarily stems from the scarcity of Verus code in existing training data, which prevents LLMs from gaining sufficient exposure to its unique verification constructs and semantics, thereby hindering their proficiency in producing correct Verus code.

%struggle to generate even syntactically correct Verus code. 

%The third challenge is that, even when the syntax is correct, LLMs lack a robust understanding of Verus's  Constructs such as ghost modes, specification-only types, and proof combinators extend beyond standard Rust and are tightly coupled with the logical discipline enforced by the verifier. As a result, the model frequently proposes specifications or invariants that are semantically incompatible with the verification rules, leading to proofs that fail to discharge.

%\paragraph{Our Approach}
To address the first challenge, we build on previous work~\cite{autoverus} by adding two new modules: (1) a \emph{View} module responsible for producing a mathematical abstraction of the data structure; and (2) a \emph{Type Invariant} module, responsible for producing type invariants.  For each component, we design dedicated system and user prompts that clearly articulate the task. These prompts include: (1) the objective of the component; (2) relevant background information to help the LLM better understand the task; (3) step-by-step instructions and the required output format; and (4) in-context learning examples~\cite{incontextlearning}.
Additionally, because not all verification components are necessary for every task, we introduce a \emph{planner} module, which analyzes the verification task as an initial step and decides which components need to be generated.

To address the second challenge, we pursued two approaches. First, we developed a suite of automated repair modules to identify and correct errors in the generated annotations. Compared with the previous work~\cite{autoverus} which focused solely on repairing proof blocks, we further introduce dedicated repair modules targeting views, type invariants, and specifications. Specifically, to facilitate the joint verification of multiple methods and tests, \toolname{} includes a repair module that refines specifications whenever the verifier reports a failed test verification. Second, we embedded structured syntax guidelines into the prompt to mitigate syntax errors in generated annotations. We sourced these guidelines from the Verus tutorial~\cite{verus-tutorial} and from standard Verus libraries~\cite{verus-stdlib}.

We evaluate \toolname{} on eleven data-structure benchmarks drawn from Verus examples and third-party code. Our framework automatically produces verified implementations for ten of these benchmarks while substantially reducing the amount of handwritten annotation required from developers.
When writing unit tests, we observe that achieving high coverage is both challenging and labor-intensive. To alleviate this burden, we apply LLM-assisted techniques to automatically generate unit test cases.

%\paragraph{Our Contribution}
To summarize, this paper makes the following key contributions:
\begin{enumerate}
    \item we introduce a new LLM-assisted workflow for generating program verification annotations for data-structure modules;
    \item we provide an implementation of our workflow in the \toolname{} tool; and
    \item we evaluate \toolname{} on eleven data structure benchmarks, demonstrating its effectiveness.
\end{enumerate}

The prompt and source code of \toolname{} is available on GitHub~\cite{artifact}.

\begin{comment}
A common approach to data structure verification involves:
\begin{enumerate}
\item implementing the data structure in a proof-oriented language (e.g., Frama-C~\cite{framac}, Dafny~\cite{dafny}, or Verus~\cite{verus}); and
\item annotating the code with specifications that describe the intended functionality of each function, along with correctness proofs demonstrating that the implementation satisfies these specifications.
\end{enumerate}
Once these steps are complete, an SMT-based verifier will check the correctness of both the specifications and the accompanying proofs. 
However, writing specifications and proofs require significant human efforts. 
\end{comment}

\begin{figure*}[htbp]
\begin{minipage}[t]{0.5\linewidth}
\begin{lstlisting}[language=Rust,style=boxed,showstringspaces=false,
firstnumber=1,
linebackgroundcolor={%
    \ifthenelse{
    \value{lstnumber} > 7 \AND \value{lstnumber} < 39
    }
    {\color{green!10}}{}
    },
]
verus!{ // Indicates the Verus environment

pub struct RingBuffer<T: Copy> {
    ring: Vec<T>,
    head: usize,
    tail: usize,
}
// The View trait, which tells the Verus verifier
// how to logically represent RingBuffer
impl<T: Copy> View for RingBuffer<T> {
    type V = (Seq<T>, usize);
    closed spec fn view(&self) -> Self::V {
        let cap = self.ring.len();
        let content = 
        if self.tail >= self.head {
            (self.ring)@.subrange(
                self.head as int, self.tail as int
            )
        } else {
            (self.ring)@.subrange(
                self.head as int, cap as int
            ).add((self.ring)@.subrange(
                0, self.tail as int
            ))
        };
        (content, cap)
    }
}
\end{lstlisting}
\end{minipage}
\begin{minipage}[t]{0.5\linewidth}
\begin{lstlisting}[language=Rust,style=boxed,showstringspaces=false,
firstnumber=28,
linebackgroundcolor={%
    \ifthenelse{
    \value{lstnumber} > 28 \AND \value{lstnumber} < 37 \OR
    \value{lstnumber} > 37 \AND \value{lstnumber} < 43 \OR
    \value{lstnumber} > 47 \AND \value{lstnumber} < 50 \OR
    \value{lstnumber} > 50 \AND \value{lstnumber} < 54
    }
    {\color{green!10}}{}
    },
]
impl RingBuffer<T> {
    // The type invariant, specifying the properties that
    // RingBuffer object must satisfy
    #[verifier::type_invariant]
    closed spec fn inv(&self) -> bool {
        &&& self.head < self.ring.len()
        &&& self.tail < self.ring.len()
        &&& self.ring.len() > 0
    }
    pub fn new(ring: Vec<T>) -> (ret: RingBuffer<T>)
    requires
        ring.len() >= 1
    ensures
        ret@.0.len() == 0,
        ret@.1 == ring.len() as nat
    {
        // Create an empty ring buffer
        RingBuffer { head: 0, tail: 0, ring }
    }
    pub fn is_full(&self) -> (ret: bool)
    ensures
        ret == (self@.0.len() == (self@.1 - 1) as nat)
    {
        proof {
            use_type_invariant(&self);
        }
        self.head == ((self.tail + 1) % self.ring.len())
    }
    pub fn enqueue(&mut self, val: T) -> (succ: bool) { ... }
    pub fn dequeue(&self) { ... }
}
fn test_ring_buffer() { /* Unit tests (omitted). */ }
fn main() { test_ring_buffer(); }
} // End of verus!
\end{lstlisting}
\end{minipage}
\caption{Verified Ring Buffer, Lines Highlighted in Green are Annotations}
\label{fig:ring-buffer-verified}
\end{figure*}

\section{Preliminaries: Verifying a Data-Structure in Verus}
\label{sec:background}

In this section, we provide background on deductive verification and Verus.  We start by introducing an example that will be used to illustrate key concepts throughout the paper.

\begin{example}[Verified Ring Buffer]
\label{ex:rb}
A ring buffer is a fixed-size data structure that treats memory as a circular array.  Ring buffers are widely used in streaming, networking, and real-time systems~\cite{tock,knuth1997art}. \Cref{fig:ring-buffer-verified} shows the Rust implementation for a ring buffer, including the Verus annotations necessary for verifying key properties of the implementation.\qed    
\end{example}

Generally speaking, deductive program verification augments programs with \emph{logical annotations}.  %Specifications state what must be true about the code.
These annotations connect directly to correctness: the verifier generates proof obligations from the code and annotations; if all obligations are proved, then the implementation is correct with respect to the annotations. Otherwise, the verifier produces an error message. Below, we describe the different categories of annotations available in the Verus program verification framework.

\paragraph{Specification Functions}
In Verus, functions are classified into two categories: \emph{specification functions} and \emph{executable functions}. Specification functions, marked with the \texttt{spec} keyword (e.g., \texttt{view} and \texttt{inv} in \Cref{fig:ring-buffer-verified}), are pure and cannot modify any global or mutable state. They are simply logical definitions of pure functions without side effects and are called exclusively from specifications, invariants, and proofs. In contrast, executable functions (e.g., \texttt{new}, \texttt{is\_full}, \texttt{enqueue}, and \texttt{dequeue}) define concrete program behavior and may modify state. When writing annotations, only specification functions can be called, since annotations must remain side-effect-free and semantically independent of execution.

\paragraph{Pre- and Postconditions}
Annotations include preconditions and postconditions for each function. The implicit program requirement is that: if a function is invoked with a program state satisfying the precondition, then it will return (or terminate) in a state satisfying the postcondition. In Verus, one uses the keywords \texttt{requires} (Lines~38--39) and \texttt{ensures} (Lines~40--41, 48--49) for specifying preconditions and postconditions, respectively. The verifier checks that every call site meets the callee's preconditions and that every function body establishes its postconditions from the assumed preconditions. 

\paragraph{Proof Blocks}
Implementation code may include \emph{proof blocks} embedded in the code.  These blocks provide hints to the verifier to help it discharge difficult proof obligations (Lines 51--53 in \Cref{fig:ring-buffer-verified}). 

Verus includes two additional constructs that are helpful when verifying data structure modules: views and type invariants.

\paragraph{View Traits}
The \emph{view trait} in Verus establishes a bridge between the concrete implementation of a data structure and its abstract mathematical representation used in specifications.
It requires implementing a \texttt{view()} function, which defines how an instance of a data structure maps to a logical object that the verifier can reason about. %This logical object is purely conceptual, it exists only during verification and allows specifications and proofs to focus on the logical behavior rather than the implementation details. 
While one could theoretically encode a data structure's state as a Cartesian product of all its fields, defining a custom view offers better abstraction, clarity, and proof efficiency, allowing specifications and proofs to focus on the logical behavior rather than the implementation details.

Once the view is implemented, specifications can refer to the logical abstraction using \texttt{a.view()} or its shorthand \texttt{a@}. Verus provides built-in specification types such as \texttt{nat}, \texttt{Seq<T>}, \texttt{Set<T>}, and \texttt{Map<K,V>} to represent natural numbers, sequences, sets, and maps, respectively. These types serve as good building blocks for implementing views.

\begin{example}
\label{ex:view-bg}
In \Cref{fig:ring-buffer-verified}, the \texttt{View} for \texttt{RingBuffer}~(Lines 8--28) abstracts the buffer as a pair (\texttt{Seq<T>}, \texttt{usize}). The first element in the pair is a mathematical sequence (\texttt{Seq<T>}) that reflects the elements currently stored in the buffer in order, starting from the element at \texttt{head} and continuing up to (but not including) the element at \texttt{tail}. The second element in the pair is the capacity of the ring buffer. To implement the view function, we slice and concatenate segments of the underlying ring according to the positions of \texttt{head} and \texttt{tail}, while hiding \texttt{head} and \texttt{tail} themselves.

This logical view abstracts the circular implementation, enabling specifications for buffer operations--like enqueue and dequeue--to be written as simple sequence transformations (e.g., appending or removing elements), without dealing with low-level index arithmetic or memory layout details.
\end{example}

\paragraph{Type Invariants}
\label{ex:invariant-bg}
%Second, we introduce the concept of \textit{type invariants}.
A \emph{type invariant} is a logical formula that all instances of a data structure must satisfy. In Verus, one declares a type invariant using the \texttt{\#[verifier::type\_invariant]} attribute. We can call \texttt{use\_type\_invariant} (Line~51) within a proof block to make invariant facts available in the current proof context.

\begin{example}
In \Cref{fig:ring-buffer-verified} (Lines 29--36), the ring buffer invariant states that \texttt{head} and \texttt{tail} are valid indices and that the ring has positive capacity.
\end{example}

\section{The \toolname{} Workflow}
\label{sec:overview}

In this section, we give an overview of the workflow used in \toolname{} using the \texttt{RingBuffer}~(\Cref{ex:rb}) as a running example.

%\paragraph{Input and Output}
The input to \toolname{} consists of two parts: (1) a user-provided Rust implementation of the code to be verified (e.g., the code in \Cref{fig:ring-buffer-verified} but \emph{without} any annotations---the green-highlighted lines would be absent), and (2) a unit-test suite that conveys typical usage patterns and rules out trivial specifications. The output is a fully annotated version of the code that passes the Verus verifier (e.g., the full \texttt{RingBuffer} code as shown in \Cref{fig:ring-buffer-verified}). Care is taken to ensure that both the code and the tests remain unchanged throughout the workflow.
%Below we detail how \toolname{} translates the original Rust implementation into a verified artifact.

\begin{figure}[t]
  \centering
  \includegraphics[width=\textwidth]{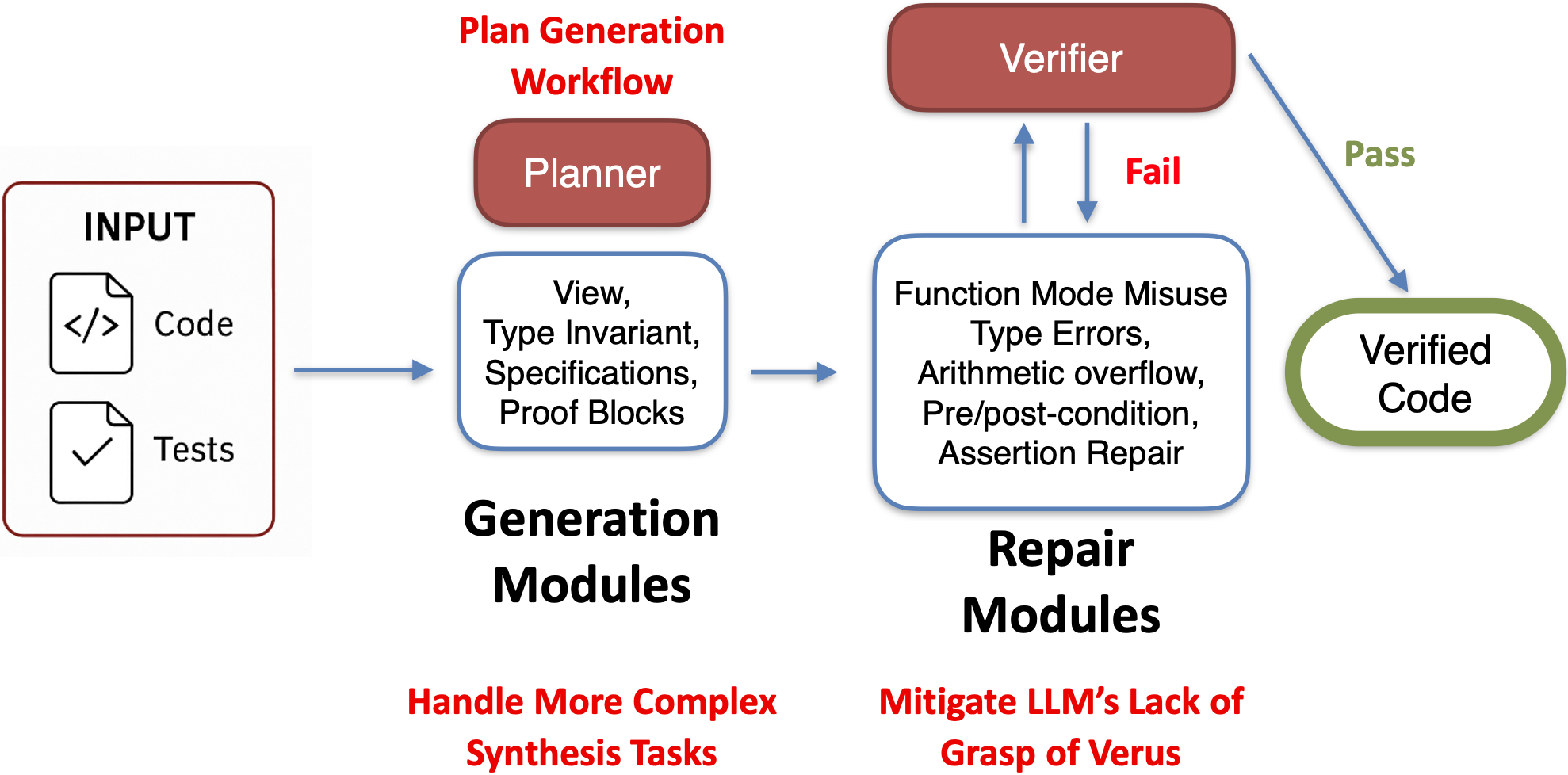}
  \caption{Workflow of \toolname{}, where rounded rectangles denote modules} %Rounded rectangles denote file artifacts (test cases, source files, verified outputs) while plain rectangles denote workflow steps; diamonds indicate decisions.
%  }
  \label{fig:workflow}
\end{figure}

The overall workflow is shown in \Cref{fig:workflow}. \toolname{} builds on the general approach of the AutoVerus system~\cite{autoverus}, in that it consists of a network of modules that collaborate to complete program verification tasks. However, as noted in the challenge discussion of \Cref{sec:intro}, verifying data structures requires techniques and components not found in AutoVerus or any prior work~\cite{autoverus,cav24llm,classinvgen,dafnysinglemethod,dafny-annotator}.

\begin{small}
\begin{algorithm}[htbp]
\SetAlgoNoLine 
\SetKwProg{Fn}{Function}{}{end}
\caption{The Outermost Algorithm of \toolname{}}
\label{fig:outmost}
\textbf{Input.} {\texttt{code}: the data structure code to be verified; \texttt{test}: unit test suite}\\
\textbf{Output.} {verified version of \texttt{code}}\\
\Fn{{{\rm \textbf{VerifyModule}}}{\rm (}{\tt code}, {\tt test}{\rm )}}{
{\tt code} $\gets$ \genfunc{}({\tt code}, {\tt test}) \textcolor{gray}{// Stage 1: Generate the Annotations}\\
\Return \fixfunc{}({\tt code}, {\tt test}) \textcolor{gray}{// Stage 2: Repair Annotations}\\
}
\end{algorithm}
\end{small}

\toolname{} applies a two-stage pipeline as shown in \Cref{fig:outmost} to synthesize annotations and perform verification.
\begin{enumerate}
\item Stage 1: Generate the initial draft of annotations through the function \genfunc{}. The detailed process is described in \Cref{sec:genfunc}.
\item Stage 2: Invoke \fixfunc{} to iteratively detect and repair incorrect annotations. The details are presented in \Cref{sec:fixfunc}.
\end{enumerate}

\noindent
Due to space limitations, we only present sketches of our prompts, but the full details of our prompts are publicly available~\cite{artifact}.

%Each stage consists of several modules with dedicated usage.

%\section{\textit{\toolname{}} in Detail}
%\label{sec:approach}

%This section provides details of \toolname{}. As mentioned above, \toolname{} takes as input a to-be-verified data-structure module \texttt{code} together with its accompanying unit test suite \texttt{test}. \toolname{} keeps the test suite unchanged, and adds annotations to the data-structure implementation \texttt{code}. Finally, \toolname{} outputs the verified version of data-structure module.

\section{Stage 1: Generating the Annotations}
\label{sec:genfunc}
\begin{small}
\begin{algorithm}[t]
\small
%\SetAlgoLined
\SetAlgoNoLine 
\SetKwProg{Fn}{Function}{}{end}
\caption{The Generation Stage}
\label{fig:genfunc}
\textbf{Parameter.} {\texttt{n}: Number of samples generated in each module invocation.}\\
\textbf{Input.} {\texttt{code}: the data structure code to be verified; \texttt{test}: unit test suite}\\
\textbf{Output.} {\texttt{code} with annotations generated}\\
\Fn{{{\rm \genfunc{}}}{\rm (}{\tt code}, {\tt test}{\rm )}}{
\texttt{plan} $\gets$ {\textbf{ExecPlanner}}(\texttt{code})\\
\ForEach{$M_i$ $\in$ {\tt plan}}{
    \texttt{code} $\gets$ {\textbf{ExecWithSampling}}($M_i$, {\tt code}, {\tt test}, {\tt n})
}
\Return {\tt code}\\
}
\end{algorithm}
\end{small}

%This section details the first stage---annotation generation. As outlined in \Cref{sec:overview}, this process involves producing four categories of annotations: ($M_1$) the \texttt{View} trait, ($M_2$) the type invariant, ($M_3$) the specifications (pre- and post-conditions), and ($M_4$) a proof blocks module for generating proof blocks and loop invariants. 
%\subsection{The Generation Stage}

The generation stage creates annotations. As mentioned above, we may need to generate different kinds of annotations, so \toolname{} uses four dedicated modules, one for each type of annotation: ($M_1$) a view module, ($M_2$) a {type invariant} module, ($M_3$) a specifications module (for specification functions, preconditions, and postconditions), and ($M_4$) a {proof blocks} module, which generates both proof blocks and invariants.

%Each category $M_i$ ($i \in {1, \dots, 4}$) is generated by a dedicated module $G_i$, responsible for producing all annotations within its respective category.%\toolname{} provides a dedicated generation module for each generating each of these. %Calling a module will generate all annotations corresponding to the invoked module.

By examining the dependencies among these kinds of annotations, we observe that it should typically be sufficient to invoke the modules in order: $\langle M_1, M_2, M_3, M_4\rangle$. At the same time, it is not always necessary to execute every module. For example, some data structures do not require non-trivial type invariants; in such cases, generating an invariant may incur: (1) more instability in the language model (and thus affect the robustness of our framework); (2) unnecessary computational cost, and (3) more runtime overhead. To mitigate these issues, we employ a \emph{planner} module that determines which generation modules should be executed.

\paragraph{Planner} The generation process begins with the planner agent (Line 5). The function \textbf{ExecPlanner} accepts the target code as input, invokes the planner agent, and outputs the modules to execute.
The planner agent instructs the LLM to select only the necessary generation modules to be executed. Its prompt comprises four blocks: (i) a concise description of the role of the planner; (ii) a catalog of the available generation modules: there are four in our current framework; (iii) background knowledge on Verus annotation generation; and (iv) a machine-parseable output format for the execution sequence.

%As discussed in \Cref{sec:overview}, dependencies among $G_i$'s permit a canonical order: $G_1 \rightarrow G_2 \rightarrow G_3 \rightarrow G_4$, but we can skip some unnecessary $G_i$'s. Thus, the planner agent instructs the LLM to determine whether each module should be executed or omitted, producing a subchain of this sequence as the final execution plan. 

The prompt is designed with several guiding principles:
\begin{enumerate}
    \item Invoke $M_1$ if the data structure can be represented by a sequence, set, or map.
    \item Invoke $M_2$ when non-trivial relationships exist among fields (e.g., range constraints or arithmetic relations).
    \item If $M_2$ is invoked, also execute $M_4$ to add type invariants into the proof context.
\end{enumerate}

%\paragraph{Planner}
 %The full details are presented in \Cref{sec:genfunc}.
%Given a to-be-verified data structure, the planner agent instructs the LLM to select and order the generation modules. Note that it is 

%Its prompt comprises four blocks: (i) a concise role description of the planner, (ii) a catalog of the available generation modules—one per annotation artifact discussed above, (iii) planning constraints that eliminate invalid orders by enforcing dependencies (e.g., generate the view and type invariant before method specifications; schedule proof blocks last), and (iv) a machine-parseable output format to ensure deterministic downstream execution. The full details are presented in \Cref{sec:approach}.

\noindent
For the ring buffer, the planner emits the full execution sequence, namely $\langle M_1, M_2, M_3, M_4\rangle$. 

%With this choice made, we next describe how \toolname{} synthesizes each component, beginning with the view generation; the remaining three components follow an analogous generation procedure.

\paragraph{Module Invocation} Following the generated plan, the framework sequentially invokes each module $M_i$ (Lines 6–10). The function \textbf{ExecWithSampling} takes as input: (1) a module to be executed; (2) the code to be verified; and (3) the unit test suite.  It executes the module, and returns the annotated code. To improve robustness, it asks for \texttt{n} samples during each module invocation~\cite{sampling}. It then selects the result yielding the maximum number of successfully verified functions.  We describe the view generation module in detail below, and then describe some differentiating aspects of the other modules, which are otherwise similar.

\paragraph{View Generation (and Refinement) Module}
To generate a view, our prompt (\Cref{fig:prompt-view-gen}) is structured as follows:
\begin{enumerate}
\item \emph{Objective.} This part articulates the task of generating a \texttt{View} implementation.%—to synthesize a minimal and semantically meaningful.
\item \emph{Verus Guidelines.}  This part provides conceptual foundations for generating a valid \texttt{View} implementation.
    As discussed in \Cref{sec:intro}, LLMs often struggle with Verus' syntax and verification-specific semantics. To mitigate this, our framework incorporates guidelines on Verus syntax and semantics imported from the Verus standard library documentation~\cite{verus-stdlib} and the official Verus tutorial~\cite{verus-tutorial}. For each different module in \toolname{}, a specific set of guidelines is added to the prompt. For \texttt{View} generation, for example, we include:
    \begin{itemize}
    \item An overview of the \texttt{View} trait (as introduced in \Cref{sec:background}).  
    \item Guidelines on how to manipulate Verus' logical types, which serve as foundational building blocks for providing a \texttt{View} implementation.  
    \item A chapter on Verus' specialized logical syntax.%, covering key constructs for writing specifications, such as using \texttt{==>} for implication.% and \texttt{&&&}/\texttt{|||} for conjunction/disjunction—to ensure syntactic validity in generated logical expressions.
    \end{itemize}

    \item \emph{Step-by-Step Instructions.} This part provides step-by-step procedural instructions on how to systematically generate a view.
    \item \emph{Examples}~\cite{incontextlearning}. This part provides examples of \texttt{View} implementations for other verified data structures (e.g., doubly-linked lists). These examples are kept the same across (and are disjoint from) all benchmarks.% and serve as reference points for few-shot in-context learning~\cite{incontextlearning}.  
    \item \emph{Code with Tests.} Finally, the prompt adds the code to be verified code along with the unit test suite.
    %This part gives examples of the view of other data-structure modules~\cite{incontextlearning}, we keep this part the same for all benchmarks.
\end{enumerate}

%Notebly, As point out in the challenge paragraph of \Cref{sec:intro}, LLM does not grasp the Verus syntax. To alliveate this isseue, we also add syntax guidance. The syntax guidance are built in chapters. We construct each chapter from the Verus standard library~\cite{?} and the Verus tutorial~\cite{?}. For example, we have syntax guidelines in manipulating Verus' axoiomic logical types~(\texttt{Seq<T>}, \texttt{Set<T>}, and \texttt{Map<K, V>}), we also have syntax guidelines in Verus' featured syntax in writing logical expressions~(e.g., use \texttt{==>} instead of the standard \texttt{->} for logical implication), these chapters will be added while generating the \texttt{View} trait, since the \texttt{View} trait requires building a view function which is some logical expression that built on top of in Verus' built-in logical types.

\begin{figure}[htbp]
\begin{lstlisting}[basicstyle=\fontsize{7pt}{7pt}\normalfont\ttfamily,frame=single,breaklines=true,escapechar=@,breakindent=0pt,xrightmargin=3em,xleftmargin=3em]
@\textbf{{[Objective]}}@
You are an expert in Verus (verifier for rust). Your task is to generate a View trait for the given data-structure module...
@\textbf{{[Verus Guidelines]}}@
Background on View and relevant Verus features...
@\textbf{{[Step-by-Step Instructions]}}@
1. Infer what should be contained in the View trait.
2. Generate the view based on the inferred information.
@\textbf{{[Examples]}}@
The View trait for other data structures, e.g., doubly-linked list.
@\textbf{{[Code and Tests]}}@
Code to be generated a View trait...
\end{lstlisting}
\caption{The Prompt for View Generation}
\label{fig:prompt-view-gen}
\end{figure}

\noindent
For the RingBuffer example, LLMs sometimes produce a trivial Cartesian-product view, which includes all of the data fields in RingBuffer (see \Cref{fig:rb-viewgen}), i.e., the view function simply maps the buffer content \texttt{self.ring} to its logical abstraction \texttt{self.ring.view()} and converts \texttt{head} and \texttt{tail} into logically defined natural numbers. 
This view function, while syntactically valid, fails to abstract away the circular structure of the ring buffer. As a result, the subsequent specification generation becomes significantly more complex: we have to explicitly reason about the arithmetic relationship between \texttt{head} and \texttt{tail} when specifying the behavior of operations such as \texttt{enqueue} and \texttt{dequeue}. %Consequently, this over-detailed view makes the specifications for these operations unnecessarily complicated and harder to verify.

\begin{figure}
\begin{lstlisting}[language=Rust,style=boxed,showstringspaces=false,basicstyle=\small\ttfamily]
impl View for RingBuffer<T> {
    type V = (Seq<T>, nat, nat);
    closed spec fn view(&self) -> Self::V {
        (self.ring.view(), head as nat, tail as nat) 
    }
}
\end{lstlisting}
\caption{Result of View Generation for RingBuffer}
\label{fig:rb-viewgen}
\end{figure}

To address this issue, we introduce an additional \emph{view refinement} step that prompts the LLM to reconsider the generated \texttt{View} and reconstruct it with fewer, more abstract components. This module encourages the model to focus on the conceptual essence of the data structure rather than on reproducing its concrete implementation. The refinement process substantially simplifies subsequent specification and proof generation tasks.

The \emph{view refinement} prompt follows the same structure as \Cref{fig:prompt-view-gen}, but emphasizes abstraction, minimality, and logical coherence over completeness. After applying this refinement, the model typically is able to produce an improved \texttt{View} implementation, like the one shown in \Cref{fig:ring-buffer-verified}.%, which succinctly captures the buffer’s logical content and capacity while abstracting away low-level representation details.

%Finally, we remark that the generation of the type invariant and specifications follows a similar pipeline; their detailed descriptions are presented in \Cref{sec:approach}.

\paragraph{Other Modules}
The prompts for the other modules share a similar structure to that shown in \Cref{fig:prompt-view-gen}. Each module prompt begins with a task description, followed by module-specific Verus guidelines.  These guidelines are specialized to the individual modules.

The \textit{Type Invariant Module} is focused on generating type invariants, which were introduced in \Cref{sec:background}.  In this module, the LLM is instructed to focus on common patterns such as range constraints, capacity checks, and arithmetic relationships between fields.%, not only we describe how type invariant works as in \Cref{sec:background}, but also instructs LLM to focus on common usages of type invariant, i.e., range constraints, capacity check, or field value arithematic bounds.

The \textit{Specification Module} is focused on generating specification functions, preconditions and postconditions.  We add mutability guidelines to ensure compatibility with Rust’s ownership and mutability system in \texttt{requires} and \texttt{ensures} clauses. For example, the LLM is prompted to use \texttt{old($\cdot$)} in \texttt{requires} to reference a variable's value prior to function invocation. %we will also add mutability guidelines, to tell LLM how to make Verus specifications compatible with Rust mutability type system in the \texttt{requires} and \texttt{ensures}. For instance, the LLM should write \texttt{old($\cdot$)} in \texttt{requires} to show that this refers to the variable before invoking this function.

The \textit{Proof Block Module} handles proof blocks and loop invariants.  The LLM is directed to construct the appropriate proof blocks and loop invariants, while explicitly incorporating type invariants into the proof context or applying relevant lemmas when necessary. %we will ask LLM do not forget to add type invariant into the context by calling \texttt{use\_type\_invariant}. 

\section{Stage 2: Repairing the Annotations}
\label{sec:fixfunc}

Unfortunately, a single generation run rarely yields fully verified code. LLMs often struggle with Verus' syntax and rules. For example, the following is an \emph{incorrect} specification that was generated for the \texttt{enqueue} method of the \texttt{RingBuffer} example:
\begin{lstlisting}[language=Rust,style=boxed,showstringspaces=false,basicstyle=\small\ttfamily]
pub fn enqueue(&mut self, val: T) -> (succ: bool)
ensures
    succ == !old(self).is_full(),
    ... // omitted
{ /* omitted */ }
\end{lstlisting}
The intent is reasonable: \texttt{enqueue} succeeds if and only if the original buffer is not full. However, \texttt{is\_full} is an \emph{executable} function (see Line~46 of \Cref{fig:ring-buffer-verified}) and therefore cannot be called from a specification context.
The verifier reports:
\begin{lstlisting}[basicstyle=\fontsize{8pt}{7pt}\normalfont\ttfamily,frame=single,breaklines=true,escapechar=@,breakindent=0pt]
Error: cannot call the executable function is_full in annotation.
\end{lstlisting}

%\paragraph{Repair Loop}
To handle such issues, \toolname{} executes an {iterative repair loop}. In each iteration, it: ($i$) selects an error reported by the verifier; ($ii$) applies a corresponding predefined repair module; and ($iii$) reruns verification. The process continues until either all errors are eliminated or a preset iteration budget is exhausted. \toolname{} provides a suite of repair modules, each targeting a specific class of errors. Error messages are routed to the appropriate module via lightweight pattern matching over the verifier's error message.
%Specifically, to support verification of unit tests, \toolname{} integrates a special repair module that strengthens the method specifications when the verifier identifies an assertion failure in the input tests (details in \Cref{sec:fixfunc}).

For the error reported above, \toolname{} invokes a dedicated mode-repair module, which provides instructions for fixing this mode misuse~(Figure~\ref{fig:prompt-mode-repair}). After applying this module, a new specification is generated.  For example, one possible solution (one we actually observed) is to create a specification version of the \texttt{is\_full} function and use it to replace all occurrences of \texttt{is\_full}. With this change, the error no longer occurs, and the workflow can continue.
\begin{lstlisting}[language=Rust,style=boxed,showstringspaces=false,basicstyle=\small\ttfamily]
spec fn is_full_spec(&self) -> bool { self@.0.len() == self@().1 - 1 }
\end{lstlisting}

%For example, \toolname{} has a module for repairing mode errors illustrated above. The module designs a dedicated prompt for fixing these errors. [Summarize the prompt in this place]. 

\begin{figure}[htbp]
\begin{lstlisting}[basicstyle=\fontsize{7pt}{7pt}\normalfont\ttfamily,frame=single,breaklines=true,escapechar=@,breakindent=0pt,xrightmargin=3em,xleftmargin=3em]
@\textbf{{[Objective]}}@
Fix the error for the following code. The error indicates that an executable function is called from annotations or vice versa...
@\textbf{{[Relevant Background]}}@
Background on mode and relevant Verus Features
@\textbf{{[Code with Error, and Tests]}}@ ...
@\textbf{{[Instructions]}}@ Make one of these changes:
1. Adjust the function to be compatible with the calling context
...
\end{lstlisting}
\caption{The Prompt for Repairing the Misuse of Specification and Executable Functions}
\label{fig:prompt-mode-repair}
\end{figure}

\begin{small}
\begin{algorithm}[htbp]
%\SetAlgoLined
\SetAlgoNoLine 
\SetKwProg{Fn}{Function}{}{end}
\caption{The Repair Stage}
\label{fig:fixfunc}
\textbf{Parameter.} {\texttt{n}: Number of samples generated in each module invocation.}\\
\textbf{Parameter.} {\texttt{m}: Maximum number of iterations in the repair loop.}\\
\textbf{Given.} {\texttt{rep\_modules}: A pre-defined set of repair modules.}\\
\textbf{Input.} {\texttt{code}: the code after generating annotations; \texttt{test}: unit test suite}\\
\textbf{Output.} {\texttt{code} with annotations repaired}\\
\Fn{{{\rm \fixfunc{}}}{\rm (}{\tt code}, {\tt test}{\rm )}}{
\For{{\tt r} $\gets$ {\tt 1} \KwTo {\tt m}}{
    {\tt err} $\gets$ {\bf TryVerify}{\rm (}{\tt code}, {\tt test}{\rm )}\\
    \If{{\tt err is None}}{\Return {\tt Success, code}}{}
    {\tt cur\_module} $\gets$ {\tt default\_module}\\
    \ForEach{{\rm (}{\tt pattern}, {\tt module}{\rm )} $\in$ {\tt rep\_modules}}{
    \If{{\tt err} {\rm matches} {\tt pattern}}{
        {\tt cur\_module} $\gets$ {\tt module}\\
        \textbf{break}
    }{}
    }
    {\tt code} $\gets$ {\textbf{ExecWithSampling}}({\tt cur\_module}, {\tt code}, {\tt test}, {\tt n})
}
\Return {\tt Failure, code}
}
\end{algorithm}
\end{small}

\Cref{fig:fixfunc} shows the repair algorithm in detail.  \toolname{} provides a predefined set of repair modules, denoted as \texttt{rep\_modules}. Each module is associated with a regular expression \texttt{pattern} specifying the class of error messages that it addresses. In the current implementation, \toolname{} includes repair modules for: (1) misuse between specification and executable functions (as demonstrated above), (2) inconsistencies with Rust’s mutability type system, (3) violations of preconditions and postconditions, (4) arithmetic overflow or underflow, (5) mismatches between logical and Rust types, (6) test assertion failures, and (7) a fallback module, \texttt{default\_module}, which prompts the LLM to attempt a repair based directly on the provided error message.

Note that the repair module for test assertion failures requires the LLM to do interprocedural analysis, something largely absent from previous work.  The repair module inspects the failing assertion, identifies the method invoked immediately before it, and then instructs the LLM to strengthen that method's postcondition so as to ensure the assertion holds\ifarxiv{} (Table~\ref{tab:repair-heuristics})\fi.

%\livia{Shan commented that this interprocedual generation should be highlighted
%}

The repair process proceeds iteratively, with a maximum number of iterations defined by the parameter \texttt{m}. In each iteration, the system first attempts to verify the current implementation via \textbf{TryVerify} (Line 8), which takes the code and test suite as input and returns the highest priority error message \texttt{err} produced by Verus. If no error is reported, verification succeeds and the code is returned (Lines 9--10). Otherwise, \texttt{err} is matched against all module patterns; if a match is found, the corresponding repair module is executed (Lines 13--15). If no match is found, the \texttt{default\_module} is invoked. Each module execution includes {\tt n} sampling attempts to improve robustness~(Line 16). %In addition to syntax-level fixes, some repair modules operate semantically—for instance, the \emph{repair test assertion failure} heuristic inspects the test assertion that triggered an error, locates the method invoked immediately before the failing assertion, and strengthens that method's postcondition so that the assertion can hold. 
If the implementation remains unverified after \texttt{m} iterations, the system reports failure and returns the final unverified result~(Line 17).

%(1) Rust's mutability type system, (2) misuse between specification and executable functions, (3) pre- and post-condition failures, (4) arithmetic overflow, (5) misuse between logical types and Rust types, and a last resort module \texttt{default\_module} which simply ask LLM to repair the error based on the provided message.

%The repair stage is performed in iterations, where the maximum number of iterations is a pre-defined parameter {\tt m}. In each iteration, we first try to verify the current implementation {\tt code} (Line 8). The function {\bf TryVerify} takes the code and the test suite as the input, and outputs the \emph{first} error message {\tt err} reported by Verus. If there is no error message, which means the code has been verified, we return the {\tt code} (Lines 9--10). Otherwise, we match the first error {\tt err} with the patterns of all repairing modules, if any of them mathches, then execute the corresponding module (Lines 13--15), if neither matches, perform the default repair  \texttt{default\_module}. The execution of each module is also coupled with sampling {\tt n} times.

%If the code is still unverified after {\tt m} rounds, report failure.
%\input{framework}
%\input{implementation}
\section{Evaluation}
\label{sec:evaluation}

In this section, we describe an evaluation of \toolname{} on a set of Rust benchmarks.

\paragraph{Benchmarks}
We evaluate \toolname{} on a benchmark set comprising eleven Rust data structure modules drawn from the Verus GitHub repository~\cite{verus} as well as other open-source repositories.
A list of the benchmarks is shown in \Cref{tab:benchmarks}.
Although the original sources are publicly available, our benchmarks have been substantially modified with more complete specifications, additional methods, and unit tests. We verified that the versions used in our evaluation do not appear verbatim in the \texttt{o1-2024-12-17} training snapshot; for example, no form of our fully verified \textsc{RingBuffer} appears in the dataset. Furthermore, even if older versions existed in pretraining data, our baseline results demonstrate that LLMs still struggle to synthesize correct Verus specifications without structured guidance---underscoring the necessity of \toolname{}'s workflow.
The benchmarks cover a set of commonly-used data structures, including a ring buffer, a vector implementation with set abstractions and common algorithms, a polymorphic option type, tree-based modules for standalone nodes and a treemap, a bitmap with 64-bit blocks, and concurrent data structures such as a read–write lock and an invariant framework for shared state. Each benchmark contains 5--21 functions to verify, where the number of functions refers to the total number of member methods in the data structure module and the corresponding test functions. In total, our benchmark set includes 129 functions to be verified across all modules. 

\begin{table*}[t]
  % \vspace*{-7.9em}

\caption{Benchmark set consisting of eleven data structures.}
\label{tab:benchmarks}
\begin{small}
\begin{tabular}{|c|p{7.8cm}|c|}
\hline
\textbf{Benchmark} & \textbf{Description} & \textbf{\#Functions} \\
\hline
\textsc{Atomics} & An atomic counter in concurrent programming. & 11 \\
\hline
\textsc{Bitmap} & Bitmap implemented over 64-bit words. & 14 \\
\hline
\textsc{Treemap} & BST-based map with ordered key/value bindings. & 21 \\
\hline
\textsc{Invariants} & Concurrent routines showing reusable invariants. & 7 \\
\hline
\textsc{Node} & Implementing the node class of the binary search tree. & 12 \\
\hline
\textsc{Option} & Polymorphic option wrapper with safe accessor methods. & 15 \\
\hline
\textsc{RingBuffer} & Implementing a ring buffer. & 13 \\
\hline
\textsc{RwLockVstd} & Read/write lock implementation. & 5 \\
\hline
\textsc{SetFromVec} & Set abstraction constructed from backing vectors. & 10 \\
\hline
\textsc{Transfer} & Account transfer routines that enforce balanced updates. & 5 \\
\hline
\textsc{Vectors} & Implementing a vector with basic algorithms. & 16 \\
\hline
\end{tabular}
\end{small}
\end{table*}
%\paragraph{Configuration}

\paragraph{Baseline}
To demonstrate the effectiveness of \toolname{}, we compare it against a baseline that iteratively invokes an LLM to generate annotations, without employing the systematic generation-and-repair workflow described in \Cref{sec:overview}. In each iteration, the baseline calls a simple \texttt{BaselineModule} that invokes the LLM once to generate all annotations simultaneously.
If the Verus verifier reports an error, we record the failing code and the associated error message, which are then injected into the prompt for subsequent iterations.
The baseline prompt includes: (1) the code to be verified, (2) the unit test suite, and (3) the previously failing code along with any corresponding error message.

We also compare against Claude Code~\cite{claudecode}, a state-of-the-art coding agent that can autonomously invoke external tools such as the Verus verifier. We configure Claude Code (using Claude Sonnet 4.5) to iteratively generate annotations and respond to Verus error messages in an agentic loop, analogous to our baseline but with the added capability of tool invocation and autonomous iteration.
%To demonstrate the effectiveness of \toolname{}, we compare against a baseline that iteratively invokes an LLM to generate annotations without any systematic generation-and-repair workflow in \Cref{sec:approach}. Specifically, in each iteration the baseline calls a simple \texttt{BaselineModule} that generates all annotations at once with the prompt consists of: (1) the code to be verified, (2) the unit test suite, and (3) previous failing code and the corresponding error message. To align with \toolname{}'s implementation, we sample \texttt{n}=3 times per invocation and keep the sample that verifies the largest number of functions. If the Verus verifier reports an error, we record the failing code and the corresponding error message and inject both into the prompt for subsequent iterations.

\paragraph{Procedure}
We run \toolname{} and the baseline on every benchmark individually. For each benchmark we first execute \toolname{} to completion and then record how many LLM invocations it required. We observe that the maximum number of LLM invocations for \toolname{} is 13, so we run the baseline on each benchmark with 13 LLM invocations. All experiments are conducted on a server running Ubuntu 22.04.5 LTS with a 24-core Intel(R) Core(TM) i9-12900K CPU and 64\,GB RAM.  We use OpenAI o1~\cite{o1} as the  back-end model for both \toolname{} and the baseline.
We instantiate \toolname{} with \texttt{n=3} and \texttt{m=5}; that is, the number of samples generated per module is 3, and the maximum number of repair rounds is set to 5. 

%Because the longest \toolname{} run requires thirteen calls, we truncate both curves at that point for visualization, ensuring that within the plotted window the baseline has at least as many calls as the pipeline even though it may issue additional calls beyond the displayed range.
%Then, we run baseline with the same number of module invocations.

\setlength{\tabcolsep}{8pt}
\begin{table}[t]
  \centering
  \caption{The Comparison between \toolname{} and Baseline.}
  \label{tab:workflow-vs-baseline}
  \begin{small}
    \begin{tabular}{cccc}
      \toprule
       & \textbf{\#Solved} & \textbf{\#Functions} \\
      \midrule
      \toolname{} & \textbf{10~$(\bf \uparrow 150.0\%)$} & \textbf{128~$(\bf \uparrow 146.2\%)$}  \\
      Claude Code & 8 & 102 \\
      {\sc Baseline} & 4 & 52 \\
      \bottomrule
    \end{tabular}
  \end{small}
\end{table}

\paragraph{Results}
We summarize the results in \Cref{tab:workflow-vs-baseline,tab:benchmark-stats}. \toolname{} successfully solves 10 out of 11 benchmarks. For the only unsolved benchmark (\textsc{Node}), \toolname{} still verifies 11 out of 12 functions. In contrast, the baseline solves only 4 out of 11 benchmarks. In terms of the total number of functions verified, \toolname{} verifies 128 out of 129 (99.2\%) functions, whereas the baseline verifies only 52. Claude Code, despite using a more recent model (Claude Sonnet 4.5) and having the ability to autonomously invoke Verus, verifies 102 functions across 8 benchmarks---substantially better than the single-prompt baseline but still short of \toolname{}'s performance. Notably, Claude Code consumes approximately 24k tokens per benchmark on average, while \toolname{} uses only 22k tokens, demonstrating that our structured workflow achieves better results with comparable or lower token cost. In summary, \toolname{} solves substantially more benchmarks and verifies significantly more functions, demonstrating that our approach improves the capabilities of LLMs to generate correct annotations.%, but also enables earlier convergence and more efficient verification.

\begin{figure}[t]
  \centering
  \includegraphics[width=0.95\textwidth]{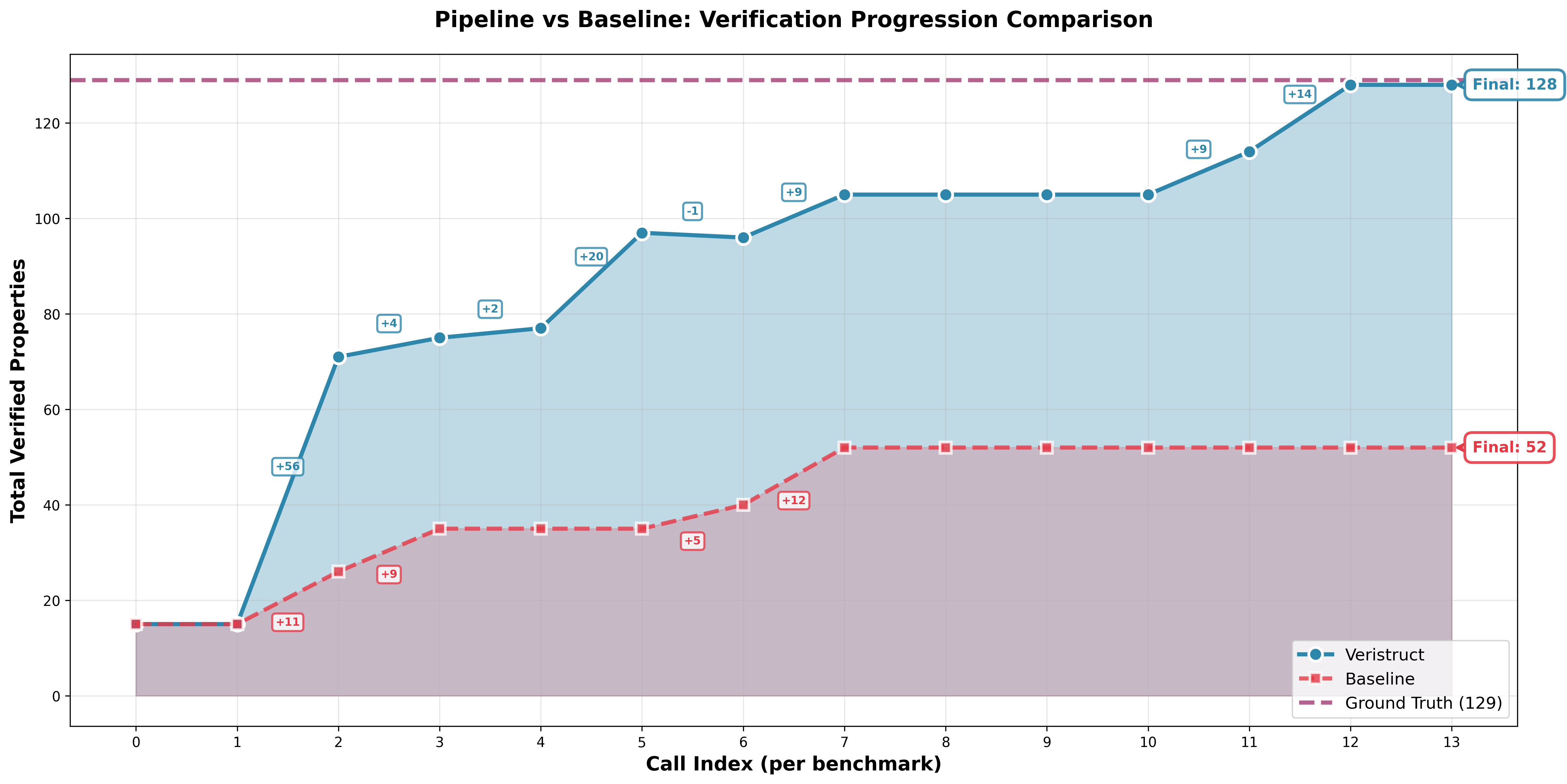}
  \caption{Aggregated verification progress across all benchmarks. \toolname{} rapidly approaches the ground-truth total of 129 functions, while the
  single-shot baseline plateaus well below full coverage.}
  \label{fig:aggregated-progress}
\end{figure}

To provide a more fine-grained comparison, we inspect the verification trajectories of \toolname{} across all benchmarks. For each integer $k$, and for each benchmark, we compute the total number of functions fully verified within the first $k$ LLM invocations.  We then sum the numbers across all benchmarks for that value of $k$, plot the result, and then increase $k$.  Plotting these points creates a visualization that shows the verification progress as a result of increasing LLM calls. We show the result in \Cref{fig:aggregated-progress}, where the x-axis denotes the LLM invocation index $k$, and the y-axis shows the cumulative number of functions verified within the first $k$ invocations across all benchmarks.

%we plot the verification progress in \Cref{fig:aggregated-progress}, where the x-axis reports the invocation index $k$ when we align per-benchmark trajectories by the $k$-th LLM call, and the y-axis denotes the cumulative number of verified functions across all benchmarks. The progress curve thus illustrates how the total number of verified functions evolves as additional calls are issued in lockstep. We also report the detailed statistics of \toolname{} in \Cref{tab:benchmark-stats}.

Although the graph primarily trends upward, there are brief dips where the number of verified functions decreases. These temporary decreases occur during specification generation, when none of the generated specifications across the requested samples is syntactically correct. In such cases, \toolname{} must adopt a candidate specification that inadvertently breaks previously verified functionality. Our pipeline deliberately retains these specifications -- even when they momentarily reduce coverage --so that subsequent repair modules can diagnose the faulty contracts once again and make progress towards verification. Typically, later stages recover from these dips and continue to progress toward full coverage.

\begin{figure*}[t]
  \centering
  \includegraphics[width=0.95\textwidth]{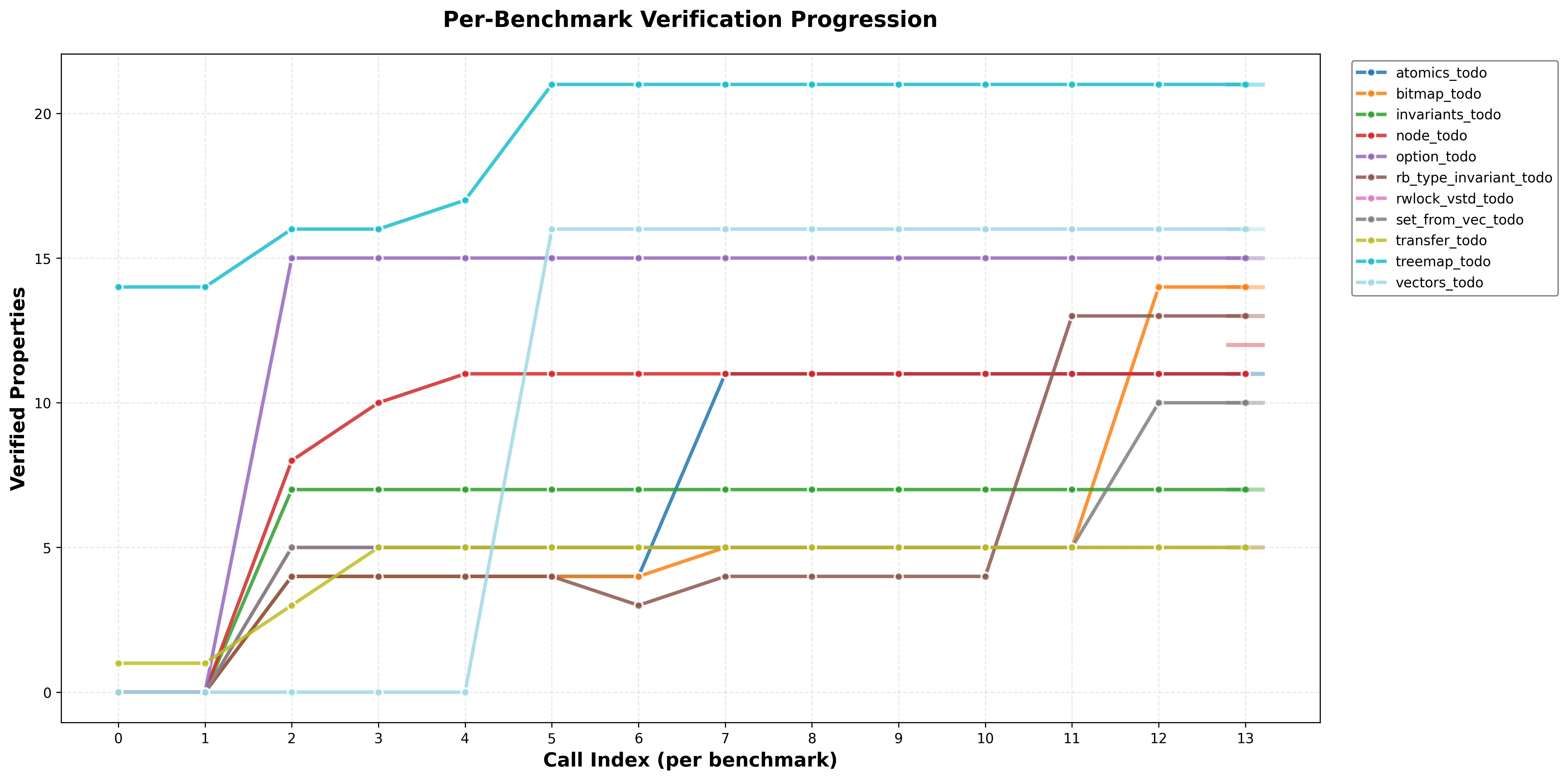}
  \caption{Per-benchmark verification progression. Each benchmark makes steady progress toward its ground-truth target through iterative refinement, demonstrating consistent improvement across diverse verification tasks. }
  \label{fig:workflow-timeline}
\end{figure*}

We can observe the consistent dominance of \toolname{} over the baseline throughout the verification process. In particular, \toolname{} achieves the largest gain in verified functions during the second invocation (the first is for the planner), highlighting the effectiveness of our planner in selecting a highly effective initial module that verifies a large portion of functions early on.
Overall, the experimental results clearly demonstrate that \toolname{} substantially outperforms the baseline, validating the effectiveness of our workflow design. \Cref{fig:workflow-timeline} further shows the per-benchmark verification progression, where each benchmark makes steady progress toward its ground-truth target through iterative refinement.%

\begin{table*}
\caption{Detailed Statistics of \toolname{}, where \#Funcs refers to the number of functions to be verified in each benchmark.}
\label{tab:benchmark-stats}
\begin{small}
\begin{center}
\begin{tabular}{|c|c|c|c|c|}
\hline
\textbf{Benchmark} & \makecell{\textbf{\#Funcs}} & \makecell{\textbf{Time}\\\textbf{~(minutes)}} & \textbf{\#LLM Calls} & \textbf{Solved?} \\
\hline
\textsc{Atomics} & 11 & 2.1 & 8 & \checkmark \\
\hline
\textsc{Bitmap} & 14 & 12.2 & 13 & \checkmark \\
\hline
\textsc{Treemap} & 21 & 0.3 & 6 & \checkmark \\
\hline
\textsc{Invariants} & 7 & 1.5 & 4 & \checkmark \\
\hline
\textsc{Node} & 12 & 10.4 & 12 & \makecell{\xmark\\(11/12 Verified)} \\
\hline
\textsc{Option} & 15 & 2.6 & 3 & \checkmark \\
\hline
\textsc{RingBuffer} & 13 & 4.2 & 12 & \checkmark \\
\hline
\textsc{RwLockVstd} & 5 & 6.4 & 3 & \checkmark \\
\hline
\textsc{SetFromVec} & 10 & 9.1 & 13 & \checkmark \\
\hline
\textsc{Transfer} & 5 & 1.2 & 4 & \checkmark \\
\hline
\textsc{Vectors} & 16 & 3.1 & 6 & \checkmark \\
\hline
\end{tabular}
\end{center}
\end{small}
\end{table*}

\paragraph{Case Study: Bitmap}
Interestingly, we find that the LLM produces a solution for the {\sc Bitmap} benchmark that is fundamentally different from the ground-truth solution. In the ground-truth implementation written by human experts, the \texttt{View} trait models the bitmap as a two-dimensional array: each \texttt{u64} block is first abstracted into a bit sequence of length 64, and then the entire bitmap data structure is represented as a 2D array of these blocks. To manipulate this representation, the ground truth defines several auxiliary functions. In contrast, the LLM adopts a simpler abstraction--it models the entire bitmap as a single array, thereby eliminating the need for auxiliary functions and reasoning directly with Verus’ built-in APIs for the \texttt{Seq} type.

%ested implementation, it first abstracts each \texttt{u64} blocks into a sequence, and then .

%handwritten view packaged the bitmap as a tuple of helper values whose structure, proposed a direct Boolean-sequence view that better matched the mental model of the data structure, which is

%The bitmap benchmark illustrates how the workflow can surface cleaner verification artifacts than the human-written baseline. The handwritten view packaged the bitmap as a tuple of helper values whose structure the authors found difficult to reproduce in LLM prompts. Our system instead proposed a direct Boolean-sequence view that better matched the mental model of the data structure. Initially we assumed this formulation would be unverifiable, but after minor adjustments elsewhere in the module the checker accepted it without additional hints. The run completes in 3.8 minutes over five iterations with six module activations (Table~\ref{tab:dynamics}), automatically discharging all fourteen functions summarized for \textsc{Bitmap} in Table~\ref{tab:benchmarks}. The final proof therefore kept the intuitive view while reducing annotation complexity, demonstrating that the workflow can validate alternatives that improve readability while preserving correctness.

\section{Related Work}
\label{sec:related}

Towards AI-assisted automated program verification, there is a long research line focusing on verifying the correctness of either an individual function~\cite{cav24llm,autoverus,dafnysinglemethod,loopy,lemur} or the type invariant for a given module~\cite{classinvgen}. In contrast, this paper targets the verification of data structure modules, which requires the synthesis of mathematical abstractions, structural invariants, and specifications and proof blocks spanning multiple methods. Consequently, none of the previous work can be applied in our setting.

%Verilib~\cite{verilib} maintains a continually expanding repository of verified Rust and Verus implementations. The open-source library aggregates formally verified modules and supporting tooling, enabling practitioners to reuse verified components rather than developing them from scratch. By lowering the barrier to accessing high-assurance software and fostering shared benchmarks for automated verification, this initiative aims to accelerate progress in LLM-assisted verification techniques, including systems such as \toolname{}.

Traditional formal-methods tools~(e.g., Frama-C~\cite{framac}, Why3~\cite{why3}, Dafny~\cite{dafny}, Coq~\cite{coq}, and Isabelle~\cite{isabelle}) provide strong correctness guarantees but require developers to manually construct detailed specifications and proofs. Thus, they impose substantial annotation overhead and demand significant expertise. In contrast, \toolname{} leverages large language models to automatically synthesize annotations. As a result, \toolname{} bridges the gap between fully manual formal verification and prior LLM-based assistants, achieving greater automation without sacrificing rigor.

\section{Conclusion and Future Work}
\label{sec:conclusion}

This paper introduced \toolname{}, a novel framework that combines large language models with Verus to automatically verify Rust data-structure modules. \toolname{} systematically generates \texttt{View} implementations, type invariants, specifications, and proof code under the scheduling of an extra planner module. To mitigate the issue that LLMs do not fully grasp Verus' annotation syntax and verification-specific semantics, we not only inject syntax guidelines in the prompt, but also introduce an additional repair stage to fix errors.%
\ifarxiv
\ Table~\ref{tab:repair-heuristics} in the appendix details the specialized heuristics that support these repairs.
\fi

We evaluated \toolname{} on eleven benchmarks drawn from the Verus GitHub repository. Across eleven benchmark modules, \toolname{} successfully verifies ten of them and verifies 128/129 (99.2\%) of functions in our benchmark. \toolname{} shows a significant improvement over the baseline, demonstrating the framework's effectiveness.

% while
%substantially reducing the handwritten annotations required compared to
%manual development. The resulting artifacts demonstrate that
%LLM-guided synthesis can substantially lower the effort required to construct
%formally verified data structures in practice.

%By decomposing the construction of proofs into generation and repair stages, the system . Furthermore, we inject syntax guidelines in the 

%This planner-guided process turns raw prover
%errors into actionable repair steps, enabling the framework to converge on
%verified implementations with limited human input; Table~\ref{tab:repair-heuristics}
%details the specialized heuristics that support these repairs.

%Looking ahead, extending the framework with richer lemma retrieval, support for
%concurrent resource algebras, and automatic unit-test generation could further
%increase its automation power. These enhancements would broaden the range of
%Rust libraries amenable to verification and bring mechanically checked
%correctness guarantees within reach for an even wider set of developers.

We conclude with several promising directions for future work.
%
%\paragraph{Toward a More Versatile Framework}
A natural next step is to extend \toolname{} to support more complex verification tasks. First, we could integrate retrieval-augmented generation (RAG)~\cite{RAG}, which could help identify useful lemmas from the Verus standard library~\cite{vstd}, strengthening proof synthesis. Second, we could additionally support the synthesis of Verus' resource algebra library~\cite{verus-practical} for verifying complex concurrent data structures. Finally, applying constrained decoding techniques~\cite{constrained-decoding} could alleviate syntax errors in generated annotations.

%Verus includes a built-in resource algebra library~\cite{verus-practical} for verifying concurrent Rust programs. By extending our LLM-assisted framework to automatically construct resource algebra objects, we can potentially enable automated verification of concurrent data structures in Rust. In addition, incorporating techniques from constrained decoding~\cite{constrained-decoding} could further reduce syntax errors and improve the overall reliability of generated proofs.

\paragraph{Automatic Unit Test Generation}
At present, our approach requires users to manually provide a comprehensive unit test suite. However, constructing such suites, especially those that achieve high coverage and include corner cases, can be both time-consuming and error-prone. Recent studies have shown that large language models are highly capable of generating high-quality test cases, including those targeting rare or edge conditions~\cite{highcovtest,liu2025llmpoweredtestcasegeneration}. Building on these advances, we plan to incorporate LLM-based unit test generation into our framework, thereby further reducing manual effort and enhancing the completeness of the verification process.

\paragraph{Reinforcement Learning for Verification}
Another promising direction is to apply reinforcement learning (RL) to specification inference and proof generation. The Verus verifier provides a natural reward signal: successful verification yields positive feedback, while failed verification attempts with diagnostic messages can guide policy updates. By training models to optimize for verification success, RL could enable more effective exploration of the specification and proof search space, potentially discovering annotation strategies that are difficult to obtain through prompting alone.

% Constraint Decoding: Control the output to ensure the 

\paragraph{Data Availability Statement}
The source code of \toolname{}, including all prompts and the benchmark suite used in our evaluation, is publicly available at \url{https://github.com/ChuyueSun/VeriStruct}.

\paragraph{Funding}
This work was funded in part by the Defense Advanced Research Projects Agency (DARPA) contract FA875024-2-1001, a gift from the Beneficial AI Foundation, and by the Stanford Center for Automated Reasoning.

\FloatBarrier
\newpage
\bibliographystyle{splncs04}
\bibliography{reference.bib}

@inproceedings{framac,
  author       = {Pascal Cuoq and Florent Kirchner and Nikolai Kosmatov and Virgile Prevosto and Julien Signoles and Boris Yakobowski},
  title        = {{Frama-C}: A Software Analysis Perspective},
  booktitle    = {Proceedings of the 10th International Conference on Software Engineering and Formal Methods (SEFM 2012)},
  editor       = {George Eleftherakis and Mike Hinchey and Mike Holcombe},
  series       = {Lecture Notes in Computer Science},
  volume       = {7504},
  pages        = {233--247},
  year         = {2012},
  publisher    = {Springer},
  address      = {Berlin, Heidelberg},
  doi          = {10.1007/978-3-642-33826-7\_16},
}

@inproceedings{dafny,
  author       = {K. Rustan M. Leino},
  title        = {{Dafny}: An Automatic Program Verifier for Functional Correctness},
  booktitle    = {Logic for Programming, Artificial Intelligence, and Reasoning (LPAR 2010)},
  series       = {Lecture Notes in Computer Science},
  volume       = {6355},
  pages        = {348--370},
  year         = {2010},
  publisher    = {Springer},
  address      = {Berlin, Heidelberg},
  doi          = {10.1007/978-3-642-17511-4\_20},
}

@article{verus,
  author       = {Andrea Lattuada and Travis Hance and Chanhee Cho and Matthias Brun and Isitha Subasinghe and Yi Zhou and Jon Howell and Bryan Parno and Chris Hawblitzel},
  title        = {{Verus}: Verifying {Rust} Programs using Linear Ghost Types},
  journal      = {Proceedings of the ACM on Programming Languages},
  year         = {2023},
  doi          = {10.1145/3586037},
}

@inproceedings{cav24llm,
  author    = {Wen, Cheng and Cao, Jialun and Su, Jie and Xu, Zhiwu and Qin, Shengchao and He, Mengda and Li, Haokun and Cheung, Shing-Chi and Tian, Cong},
  title     = {Enchanting Program Specification Synthesis by Large Language Models Using Static Analysis and Program Verification},
  booktitle = {Computer Aided Verification (CAV 2024)},
  series    = {Lecture Notes in Computer Science},
  volume    = {14682},
  pages     = {302--328},
  publisher = {Springer},
  year      = {2024},
  doi       = {10.1007/978-3-031-65630-9\_16},
}

@book{knuth1997art,
  title={The Art of Computer Programming, Volume 1: Fundamental Algorithms},
  author={Knuth, Donald E},
  year={1997},
  publisher={Addison-Wesley}
}

@inproceedings{pldi-invariant,
author = {Miltner, Anders and Padhi, Saswat and Millstein, Todd and Walker, David},
title = {Data-driven inference of representation invariants},
year = {2020},
isbn = {9781450376136},
publisher = {Association for Computing Machinery},
address = {New York, NY, USA},
url = {https://doi.org/10.1145/3385412.3385967},
doi = {10.1145/3385412.3385967},
booktitle = {Proceedings of the 41st ACM SIGPLAN Conference on Programming Language Design and Implementation},
pages = {1–15},
numpages = {15},
location = {London, UK},
series = {PLDI 2020}
}

@online{verus-stdlib,
author = {{Verus Contributors}},
title = {vstd: Verus Standard Library API Documentation},
year = {2025},
url = {https://verus-lang.github.io/verus/verusdoc/vstd/}
,
urldate = {2025-10-10},
organization = {verus-lang}
}

@online{artifact,
title = {Artifact of VeriStruct},
year = {2025},
url = {https://github.com/ChuyueSun/VeriStruct},
urldate = {2025-10-10},
}

@inproceedings{test-intent,
  title={Evaluating llm-driven user-intent formalization for verification-aware languages},
  author={Lahiri, Shuvendu K},
  booktitle={CONFERENCE ON FORMAL METHODS IN COMPUTER-AIDED DESIGN--FMCAD 2024},
  pages={142},
  year={2024}
}

@online{verus-tutorial,
author = {{Verus Contributors}},
title = {Verus Tutorial and Reference},
year = {2025},
url = {https://verus-lang.github.io/verus/guide/}
,
urldate = {2025-10-10},
organization = {verus-lang}
}

@techreport{o1,
  title        = {OpenAI o1 System Card},
  author       = {{OpenAI}},
  institution  = {{OpenAI}},
  year         = {2024},
  type         = {Technical Report},
  note         = {Includes safety evaluations and red teaming results for the o1 and o1-mini models},
}

@inproceedings{insecure1,
  author    = {Neil Perry and Megha Srivastava and Deepak Kumar and Dan Boneh},
  title     = {Do Users Write More Insecure Code with {AI} Assistants?},
  booktitle = {Proceedings of the 2023 ACM SIGSAC Conference on Computer and Communications Security (CCS)},
  year      = {2023},
  pages     = {2785--2799},
  publisher = {ACM},
  doi       = {10.1145/3576915.3623157}
}

@article{sampling,
  author       = {Sun, Chung-En and Gao, Sicun and Weng, Tsui-Wei},
  title        = {Breaking the Barrier: Enhanced Utility and Robustness in Smoothed DRL Agents},
  journal      = {ICML},  
  year         = {2024}
}

@article{constrained-decoding,
  title={Constrained decoding for secure code generation},
  author={Fu, Yanjun and Baker, Ethan and Ding, Yu and Chen, Yizheng},
  journal={arXiv preprint arXiv:2405.00218},
  year={2024}
}

@inproceedings{bug1,
  author    = {Zhijie Wang and Zijie Zhou and Da Song and Yuheng Huang and Shengmai Chen and Lei Ma and Tianyi Zhang},
  title     = {Towards Understanding the Characteristics of Code Generation Errors Made by Large Language Models},
  booktitle = {Proceedings of the 47th IEEE/ACM International Conference on Software Engineering (ICSE)},
  year      = {2025},
  note      = {To appear}
}

@article{loopy,
  title={Finding inductive loop invariants using large language models},
  author={Kamath, Adharsh and Senthilnathan, Aditya and Chakraborty, Saikat and Deligiannis, Pantazis and Lahiri, Shuvendu K and Lal, Akash and Rastogi, Aseem and Roy, Subhajit and Sharma, Rahul},
  journal={arXiv preprint arXiv:2311.07948},
  year={2023}
}

@inproceedings{lemur,
  title={Lemur: Integrating Large Language Models in Automated Program Verification},
  author={Wu, Haoze and Barrett, Clark and Narodytska, Nina},
  booktitle={The Twelfth International Conference on Learning Representations}
}

@inproceedings{verus-practical,
author = {Lattuada, Andrea and Hance, Travis and Bosamiya, Jay and Brun, Matthias and Cho, Chanhee and LeBlanc, Hayley and Srinivasan, Pranav and Achermann, Reto and Chajed, Tej and Hawblitzel, Chris and Howell, Jon and Lorch, Jacob R. and Padon, Oded and Parno, Bryan},
title = {{Verus}: A Practical Foundation for Systems Verification},
year = {2024},
isbn = {9798400712517},
publisher = {Association for Computing Machinery},
address = {New York, NY, USA},
url = {https://doi.org/10.1145/3694715.3695952},
doi = {10.1145/3694715.3695952},
booktitle = {Proceedings of the ACM SIGOPS 30th Symposium on Operating Systems Principles},
pages = {438–454},
numpages = {17},
location = {Austin, TX, USA},
series = {SOSP '24}
}

@article{dafnysinglemethod,
author = {Misu, Md Rakib Hossain and Lopes, Cristina V. and Ma, Iris and Noble, James},
title = {Towards AI-Assisted Synthesis of Verified {Dafny} Methods},
year = {2024},
issue_date = {July 2024},
publisher = {Association for Computing Machinery},
address = {New York, NY, USA},
volume = {1},
number = {FSE},
url = {https://doi.org/10.1145/3643763},
doi = {10.1145/3643763},
journal = {Proc. ACM Softw. Eng.},
month = jul,
articleno = {37},
numpages = {24}
}

@article{classinvgen,
  author  = {Sun, Chuyue and Agashe, Viraj and Chakraborty, Saikat and Taneja, Jubi and Barrett, Clark W. and Dill, David L. and Qiu, Xiaokang and Lahiri, Shuvendu K.},
  title   = {{ClassInvGen}: Class Invariant Synthesis using Large Language Models},
  journal = {CoRR},
  volume  = {abs/2502.18917},
  year    = {2025},
  url     = {https://arxiv.org/abs/2502.18917},
}

@inproceedings{clover,
  author    = {Sun, Chuyue and Sheng, Ying and Padon, Oded and Barrett, Clark},
  title     = {Clover: Closed-Loop Verifiable Code Generation},
  booktitle = {AI Verification (SAIV 2024)},
  series    = {Lecture Notes in Computer Science},
  volume    = {14846},
  pages     = {134--155},
  publisher = {Springer, Cham},
  year      = {2024},
  doi       = {10.1007/978-3-031-65112-0_7},
}

@article{incontextlearning,
  title={A survey on in-context learning},
  author={Dong, Qingxiu and Li, Lei and Dai, Damai and Zheng, Ce and Ma, Jingyuan and Li, Rui and Xia, Heming and Xu, Jingjing and Wu, Zhiyong and Liu, Tianyu and others},
  journal={arXiv preprint arXiv:2301.00234},
  year={2022}
}

@misc{vstd,
  author       = {Lattuada, Andrea and Parno, Bryan and Bosamiya, Jay and Hawblitzel, Chris and Hance, Travis and others},
  title        = {{vstd}: Verus Standard Library},
  howpublished = {\url{https://github.com/verus-lang/verus/tree/main/vstd}},
  year         = {2024},
  month        = jun,
  note         = {Version 0.0.0, MIT License},
}

@article{autoverus,
  author  = {Yang, Chenyuan and Li, Xuheng and Misu, Md Rakib Hossain and Yao, Jianan and Cui, Weidong and Gong, Yeyun and Hawblitzel, Chris and Lahiri, Shuvendu and Lorch, Jacob R. and Lu, Shuai and Yang, Fan and Zhou, Ziqiao and Lu, Shan},
  title   = {{AutoVerus}: Automated Proof Generation for {Rust} Code},
  journal = {CoRR},
  volume  = {abs/2409.13082},
  year    = {2024},
  url     = {https://arxiv.org/abs/2409.13082},
}

@inproceedings{reynolds-seplogic,
  author    = {Reynolds, John C.},
  title     = {Separation Logic: A Logic for Shared Mutable Data Structures},
  booktitle = {Proceedings of the 17th Annual IEEE Symposium on Logic in Computer Science (LICS 2002)},
  year      = {2002},
  pages     = {55--74},
  publisher = {IEEE Computer Society},
  address   = {Los Alamitos, CA, USA},
  doi       = {10.1109/LICS.2002.1029817},
}

@inproceedings{parkinson-bierman,
  author    = {Parkinson, Matthew and Bierman, Gavin},
  title     = {Separation Logic and Abstraction},
  booktitle = {Proceedings of the 32nd ACM SIGPLAN-SIGACT Symposium on Principles of Programming Languages (POPL '05)},
  year      = {2005},
  pages     = {247--258},
  publisher = {Association for Computing Machinery},
  address   = {New York, NY, USA},
  doi       = {10.1145/1040305.1040327},
}

@article{ohearn-cacm,
  author  = {O'Hearn, Peter W.},
  title   = {Separation Logic},
  journal = {Communications of the ACM},
  volume  = {62},
  number  = {2},
  pages   = {86--95},
  year    = {2019},
  month   = feb,
  doi     = {10.1145/3211968},
}

@article{dafny-annotator,
  author       = {Poesia, Gabriel and Loughridge, Chloe and Amin, Nada},
  title        = {dafny-annotator: {AI}-Assisted Verification of {Dafny} Programs},
  journal      = {CoRR},
  volume       = {abs/2411.15143},
  year         = {2024},
  url          = {https://arxiv.org/abs/2411.15143},
  note         = {arXiv:2411.15143},
}

@inproceedings{RAG,
author = {Lewis, Patrick and Perez, Ethan and Piktus, Aleksandra and Petroni, Fabio and Karpukhin, Vladimir and Goyal, Naman and K\"{u}ttler, Heinrich and Lewis, Mike and Yih, Wen-tau and Rockt\"{a}schel, Tim and Riedel, Sebastian and Kiela, Douwe},
title = {Retrieval-augmented generation for knowledge-intensive NLP tasks},
year = {2020},
isbn = {9781713829546},
publisher = {Curran Associates Inc.},
address = {Red Hook, NY, USA},
booktitle = {Proceedings of the 34th International Conference on Neural Information Processing Systems},
articleno = {793},
numpages = {16},
location = {Vancouver, BC, Canada},
series = {NIPS '20}
}

@inproceedings{highcovtest,
author = {Wang, Zejun and Liu, Kaibo and Li, Ge and Jin, Zhi},
title = {HITS: High-coverage {LLM}-based Unit Test Generation via Method Slicing},
year = {2024},
isbn = {9798400712487},
publisher = {Association for Computing Machinery},
address = {New York, NY, USA},
url = {https://doi.org/10.1145/3691620.3695501},
doi = {10.1145/3691620.3695501},
booktitle = {Proceedings of the 39th IEEE/ACM International Conference on Automated Software Engineering},
pages = {1258–1268},
numpages = {11},
location = {Sacramento, CA, USA},
series = {ASE '24}
}

@misc{liu2025llmpoweredtestcasegeneration,
      title={LLM-Powered Test Case Generation for Detecting Bugs in Plausible Programs}, 
      author={Kaibo Liu and Zhenpeng Chen and Yiyang Liu and Jie M. Zhang and Mark Harman and Yudong Han and Yun Ma and Yihong Dong and Ge Li and Gang Huang},
      year={2025},
      eprint={2404.10304},
      archivePrefix={arXiv},
      primaryClass={cs.SE},
      url={https://arxiv.org/abs/2404.10304}, 
}

@inproceedings{tock,
author = {Levy, Amit and Campbell, Bradford and Ghena, Branden and Giffin, Daniel B. and Pannuto, Pat and Dutta, Prabal and Levis, Philip},
title = {Multiprogramming a 64kB Computer Safely and Efficiently},
year = {2017},
isbn = {9781450350853},
publisher = {Association for Computing Machinery},
address = {New York, NY, USA},
url = {https://doi.org/10.1145/3132747.3132786},
doi = {10.1145/3132747.3132786},
booktitle = {Proceedings of the 26th Symposium on Operating Systems Principles},
pages = {234–251},
numpages = {18},
location = {Shanghai, China},
series = {SOSP '17}
}

@misc{coq,
  title={The {Coq} proof assistant},
  author={{The Coq Development Team}},
  year={2024},
  howpublished={\url{https://coq.inria.fr}},
  note={Version 8.19.0}
}

@misc{isabelle,
  title={Isabelle},
  author={Nipkow, Tobias and Paulson, Lawrence C. and Wenzel, Makarius},
  year={2024},
  howpublished={\url{https://isabelle.in.tum.de}},
  note={Version 2023}
}

@misc{why3,
  title={Why3},
  author={Filliâtre, Jean-Christophe and Paskevich, Andrei},
  year={2024},
  howpublished={\url{https://why3.lri.fr}},
  note={Version 1.6.0}
}

@misc{verilib,
  title        = {Verilib},
  author       = {{Verus Project}},
  howpublished = {\url{https://verilib.org/}},
  note         = {Accessed: 2025-02-17},
}

@article{BBB+23,
   author = {Clark Barrett and Brad Boyd and Elie Bursztein and Nicholas
	Carlini and Brad Chen and Jihye Choi and Amrita Roy Chowdhury and
	Mihai Christodorescu and Anupam Datta and Soheil Feizi and
	Kathleen Fisher and Tatsunori Hashimoto and Dan Hendrycks and
	Somesh Jha and Daniel Kang and Florian Kerschbaum and Eric
	Mitchell and John Mitchell and Zulfikar Ramzan and Khawaja Shams
	and Dawn Song and Ankur Taly and Diyi Yang},
   title = {Identifying and Mitigating the Security Risks of Generative AI},
   journal = {Foundations and Trends in Privacy and Security},
   volume = {6},
   number = {1},
   pages = {1--52},
   publisher = {now publishers inc.},
   year = {2023},
   issn = {2474-1558},
   doi = {10.1561/3300000041},
   url = {http://dx.doi.org/10.1561/3300000041}
}

@incollection{BSST21,
   author = {Clark Barrett and Roberto Sebastiani and Sanjit Seshia and
	Cesare Tinelli},
   editor = {Armin Biere and Marijn J. H. Heule and Hans van Maaren and
	Toby Walsh},
   title = {Satisfiability Modulo Theories},
   booktitle = {Handbook of Satisfiability, Second Edition},
   series = {Frontiers in Artificial Intelligence and Applications},
   volume = {336},
   chapter = {33},
   pages = {825--885},
   publisher = {IOS Press},
   month = feb,
   year = {2021},
   url = {http://theory.stanford.edu/~barrett/pubs/BSST21.pdf}
}

@misc{claudecode,
  title = {Claude Code},
  author = {Anthropic},
  year = {2025},
  howpublished = {\url{https://claude.ai/claude-code}},
  note = {Accessed: 2025}
}

\newpage
\appendix
\section{Supplementary Material}
\label{appendix:supp}

\Cref{tab:repair-heuristics} summarizes the specialized repair heuristics used by \toolname{}, the failure classes they target, and representative corrective actions.

\begin{table}[h]
  \caption{Specialized repair heuristics, the failure classes they target, and representative actions.}
  \label{tab:repair-heuristics}
  \begin{small}
  \begin{tabular}{p{0.15\textwidth}p{0.30\textwidth}p{0.42\textwidth}}
    \hline
    \textit{Heuristic} & \textit{Primary failures} & \textit{Representative actions} \\
    \hline
    Syntax & Parser and sequence syntax errors & Insert missing \texttt{view()} calls or lemmas, balance delimiters, correct operators/generics \\
    Type & Type mismatches, constructor invariant failures & Rewrite expressions, add explicit generics (e.g., \texttt{None::<T>}), patch missing \texttt{requires} clauses \\
    Arithmetic & Nonlinear reasoning, overflow/underflow & Emit \texttt{by(nonlinear\_arith)} proofs, assert bounds, strengthen loop invariants \\
    Precondition & Preconditions, vector bounds, private access & Introduce proof blocks, assert length/index constraints, adjust permission-violating calls \\
    Postcondition & Failing \texttt{ensures} clauses & Add exit proofs, tune invariants, expose ghost accessors for private fields \\
    Invariant & Loop invariant failures pre-/post-loop & Assert invariants before loops, propagate conditions, revise invalid invariants \\
    Missing Element & Absent imports or trait implementations & Insert \texttt{use} statements, synthesize required trait methods with contracts \\
    Mode & Mode mismatches, visibility issues & Wrap calls in spec/proof blocks, retag functions, set open/closed attributes \\
    Old-Self & Missing \texttt{old(self)} references & Rewrite \texttt{self.} in \texttt{requires} blocks to \texttt{old(self).} \\
    Assertion & Failed assertions, test expectations & Add reveals or lemmas and lift tests into postconditions \\
    Decrease & Unsatisfied \texttt{decreases} obligations & Prove measure drops, adjust \texttt{decreases} expressions, update loop variables \\
    Inv.\ Removal & Redundant invariant invocations & Delete unnecessary \texttt{self.inv()} calls enforced by type invariant attributes \\
    \hline
  \end{tabular}
  \end{small}
\end{table}

\end{document}